\documentclass{article}
\usepackage{spconf,amsmath,graphicx}

\newcommand{\argmax}{\mathop{\rm arg~max}\limits} 

\usepackage{amssymb}
\usepackage{enumerate}
\usepackage{graphicx}
\usepackage{multirow}
\usepackage{arydshln}
\usepackage{color}
\usepackage{float}
\usepackage{bm}
\usepackage{comment}
\usepackage{cite}
\usepackage{url}
\usepackage{amsmath}
\usepackage{algorithmic}
\usepackage{array}
\usepackage{colortbl}
\usepackage{subfigure}

\newlength\savedwidth
\newcommand{\wcline}[1]{\noalign{\global\savedwidth\arrayrulewidth\global\arrayrulewidth 1.0pt} \cline{#1}
\noalign{\global\arrayrulewidth\savedwidth}}

\title{Graph Cepstrum: Spatial Feature\\Extracted from Partially Connected Microphones}
\name{Keisuke Imoto\vspace{-8pt}}
\address{Ritsumeikan University, Japan}
\begin{document}
%
\maketitle
%
\begin{abstract}
In this paper, we propose an effective and robust method of spatial feature extraction for acoustic scene analysis utilizing partially synchronized and/or closely located distributed microphones.
In the proposed method, a new cepstrum feature utilizing a graph-based basis transformation to extract spatial information from distributed microphones, while taking into account whether any pairs of microphones are synchronized and/or closely located, is introduced.
Specifically, in the proposed graph-based cepstrum, the log-amplitude of a multichannel observation is converted to a feature vector utilizing the inverse graph Fourier transform, which is a method of basis transformation of a signal on a graph.
Results of experiments using real environmental sounds show that the proposed graph-based cepstrum robustly extracts spatial information with consideration of the microphone connections.
Moreover, the results indicate that the proposed method more robustly classifies acoustic scenes than conventional spatial features when the observed sounds have a large synchronization mismatch between partially synchronized microphone groups.
%
\end{abstract}
%
\section{Introduction}
\label{sec:intro}
Analyzing scenes from sounds, which is called acoustic scene analysis (ASA) \cite{Imoto_AST2018_01}, has many applications such as systems for monitoring elderly people or infants \cite{Peng_ICME2009_01,Guyot_ICASSP2013_01}, automatic surveillance systems \cite{Harma_ICME2005_01,Radhakrishnan_WASPAA2005_01,Ntalampiras_ICASSP2009_01}, automatic file-logging systems \cite{Eronen_TASLP2006_01,Imoto_IEICE2016_01,Schroder_ICASSP2016_01}, and advanced multimedia retrieval \cite{Zhang_TASLP2001_01,Jin_INTERSPEECH2012_01,Ohishi_ICASSP2013_01,Liang_ICASSP2017_01}.

Many methods of ASA utilizing spectral information have been developed; for instance, Eronen {\it et al.} \cite{Eronen_TASLP2006_01} and Mesaros {\it et al.} \cite{Mesaros_EUSIPCO2010_01} have proposed methods based on mel-frequency cepstral coefficients (MFCCs) and Gaussian mixture models (GMMs).
Han {\it et al.} \cite{Han_DCASE2017_01} and Jallet {\it et al.} \cite{Jallet_DCASE2017_01} have proposed methods based on mel-spectrograms and the convolutional neural network (CNN).
Kim {\it et al.} \cite{Kim_WASPAA2009_01} and Imoto {\it et al.} \cite{Imoto_IEICE2016_01,Imoto_MLSP2013_01} have developed methods utilizing Bayesian generative models of acoustic word sequences.

More recently, ASA based on spatial information has also been proposed \cite{Kwon_ISCS2009_01,Giannoulis_EUSIPCO2015_01, Dekkers_DCASE2017_01,Dekkers_arXiv2018_01,Tanabe_DCASE2018_01}.
Many of these methods extract spatial information based on time differences or sound power ratios between channels; therefore, it is necessary that the microphones are synchronized between channels and the microphone locations or array geometry are known.
To synchronize the multiple microphones and determine the microphone locations or array geometry, we need multiple A/D converters controlled by a common temporal clock and a dedicated microphone array.
However, in realistic situations, this is not always feasible.

%
To extract spatial information using unsynchronized distributed microphones whose locations and array geometry are unknown, K\"{u}rby \textit{et al.} \cite{Kurby_DCASE2016_01} devised scene classification methods based on the late fusion of scene classification results obtained with each microphone.
Imoto and Ono have proposed a spatial cepstrum that can be applied under the unsynchronized and blind condition \cite{Imoto_EUSIPCO2015_01,Imoto_TASLP2017_01}.
In the extraction of a spatial cepstrum, log-power observations recorded by multiple microphones are converted to a feature vector by a basis transformation similarly to the cepstrum, except that principal component analysis (PCA) is applied for the basis transformation.

\begin{figure}[t]
\centering
\includegraphics[width=1.0\columnwidth]{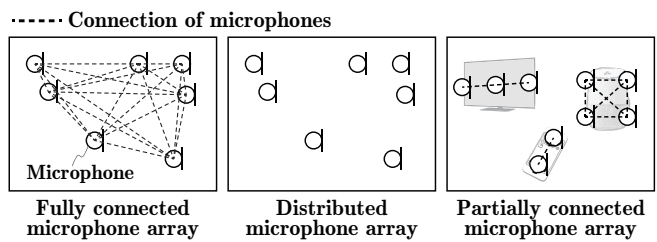}
\caption{Example of microphone connections}
\label{fig:connection01}
\end{figure}

Recently, the numbers of smartphones, smart speakers, and IoT devices that have multiple microphones have rapidly been increasing.
A microphone array composed of these multiple microphones are often partially synchronized and closely located as shown in Fig.~\ref{fig:connection01}, where we collectively refer to synchronized or closely located microphones as {\it connected} microphones.
%
%
%
The time delay and sound power ratio between channels are significant cues for extracting spatial information even when microphones are partially connected.
However, in conventional methods of spatial feature extraction using a distributed microphone array, it is not considered whether microphones are partially connected.
\par In this paper, a spatial feature extraction method for a distributed microphone array that can take into account whether any pairs of microphones are connected is proposed.
To consider whether any pairs of microphones are connected, we utilize a graph representation of the microphone connections, in which the power observations and microphone connections are represented by weights of the nodes and edges, respectively. 
Using the proposed method, we then extract spatial information by applying a graph Fourier transform, which enables spatial feature extraction considering the connections between microphones.

The rest of this paper is structured as follows.
In section 2, the cepstrum and spatial cepstrum, which are respectively used for conventional spectral and spatial feature extraction, are introduced.
In sections 3, the proposed spatial feature extraction for partially connected distributed microphones and the similarity of the proposed method to the cepstrum and spatial cepstrum are discussed.
In section 4, results of experiments using the proposed and conventional spatial features are reported.
In section 5, this paper is concluded.
%
\section{Conventional Cepstrum Features}
\label{sec:SC}
%
In this section, we discuss the cepstrum and spatial cepstrum \cite{Imoto_EUSIPCO2015_01,Imoto_TASLP2017_01}, which are the conventional spectral and spatial features, respectively.

Suppose that a multichannel observation is recorded by $N$ microphones and let $s_{\omega,\tau,n}$ be the 
short-time Fourier transform (STFT) representation of an observation for microphone $n$, where $\omega$ and $\tau$ represent the frequency and time frame, respectively.
We then designate the amplitude component of $s_{\omega,\tau,n}$ as $a_{\omega,\tau,n}=|s_{\omega,\tau,n}|$.
%
\subsection{Cepstrum as Spectral Feature}
\label{ssec:SpectralCepstrum}
Let us consider the frequency-based log-amplitude vector
%
\begin{align}
{\bf p}_{\tau} = \left(
  \begin{array}{c}
    \log \bar{a}_{1,\tau} \\
    \log \bar{a}_{2,\tau} \\
    \vdots \\
    \log \bar{a}_{\omega,\tau}\\
    \vdots \\
    \log \bar{a}_{\Omega,\tau}
  \end{array}
  \right),
\end{align}
%
\noindent where $\Omega$ is the number of frequency bins and
%
\begin{align}
\bar{a}_{\omega,\tau}=\sqrt{\frac{1}{N}\sum_n a_{\omega,\tau,n}^2}
\end{align}
%
\noindent is the root mean square (RMS) of the amplitude spectrum over the channels.
The inverse discrete Fourier transform (IDFT) of ${\bf p}_{\tau}$, which is defined as
%
\begin{align}
{\bf c}_{\tau}={\bf Z}_\Omega {\bf p}_{\tau},
\end{align}
%
\noindent is called the cepstrum, where ${\bf Z}_\Omega \in \mathbb{R}^{\Omega \times \Omega}$ is called the IDFT matrix.
The mel-frequency cepstrum coefficient (MFCC), which has been widely used as a spectral feature, is also defined using the discrete cosine transform (DCT) for a mel-frequency representation similarly to the cepstrum.
%
\subsection{Spatial Cepstrum as Spatial Feature}
\label{subsec:SC}
As an extraction technique of spatial information for unsynchronized distributed microphones whose locations and array geometry are unknown, the spatial cepstrum (SC), which is an extension of the cepstrum feature to the spatial domain feature, has been proposed \cite{Imoto_EUSIPCO2015_01,Imoto_TASLP2017_01}.

In unsynchronized distributed microphones, synchronization over channels is a challenging problem; thus, phase information, which is susceptible to synchronization mismatch, may be unreliable.
To design a more robust spatial feature to synchronization mismatch, the SC also utilizes only the log-amplitude vector
%
\begin{align}
{\bf q}_{\tau} = \left(
  \begin{array}{c}
    \log \tilde{a}_{\tau,1} \\
    \log \tilde{a}_{\tau,2} \\
    \vdots \\
    \log \tilde{a}_{\tau,n} \\
    \vdots \\
    \log \tilde{a}_{\tau,N}
  \end{array}
  \right),
\end{align}
%
\noindent where
%
\begin{align}
\tilde{a}_{\tau,n}=\sqrt{\frac{1}{\Omega}\sum_{\omega} a_{\omega,\tau,n}^2}
\end{align}
%
\noindent is the RMS of the amplitude spectrum over the frequency bins at each time frame.
In the case of the cepstrum feature, since $\tilde{a}_{\omega,\tau}$ represents the amplitude at each subband, which is uniformly spaced on the linear frequency or mel-frequency axis, the basis transformation using the IDFT can be applied.
However, when the distributed microphones are non-uniformly located, we cannot apply the basis transformation such as the IDFT or DCT; thus, PCA is applied to extract the SC.

Let ${\bf R}_q$ be the covariance matrix of ${\bf q}_\tau$ given by
%
\begin{align}
{\bf R}_q = \frac{1}{T}\sum_\tau {\bf q}_\tau{\bf q}_\tau^{\mathsf{T}},
\end{align}
%
\noindent where $T$ is the number of time frames.
Since ${\bf R}_{q}$ is a symmetric matrix, the eigendecomposition of ${\bf R}_{q}$ can be represented as
%
\begin{align}
{\bf R}_{q} &= {\bf E} {\bf D} {\bf E}^{\mathsf{T}},
\end{align}
%
\noindent where ${\bf E}$ and ${\bf D}$ are the eigenvector matrix and the diagonal matrix whose diagonal elements are equal to the eigenvalues in descending order, respectively.
Using this eigenvector matrix ${\bf E}$, the SC is defined as
%
\begin{align}
{\bf d}_{\tau} = {\bf E}^{\mathsf{T}} {\bf q}_{\tau}.
  \label{eq:def_SC}
\end{align}
%
The SC can extract spatial information without information on microphone locations or the array geometry, although it requires training with sound data to estimate the eigenvector matrix ${\bf E}$ by PCA.
Moreover, since the SC does not consider whether any pairs of microphones are synchronized or closely located, prior information about microphone connections or partial location information of the microphones cannot be utilized for spatial feature extraction.
%
\section{Spatial Feature Extraction Based on Graph Cepstrum}
\label{sec:GFTceps}
Consider the situation that a microphone array is composed of multiple generic acoustic sensors mounted on smartphones, smart speakers, or IoT devices, where some of the microphones mounted on each device are connected.
To extract spatial information from these microphones, methods of blind synchronization between devices can be applied \cite{Hasegawa_LVAICA2010_01,Ono_WASPAA2009_01,Schmalenstroeer_EUSIPCO2013_01,Miyabe_ESP2015_01}.
However, it is still not easy to synchronize these microphones accurately.
Moreover, synchronization is not always satisfactory because recorded sounds are easily affected by background noise, reverberation, and reflection.
The use of the SC has been proposed to address these difficulties; that is, the SC can robustly extract spatial information using a distributed microphone array.
However, the SC cannot extract spatial information considering whether microphones are partially connected.
In this paper, we propose a new method of spatial information extraction taking this advantage of the SC and taking partial connections of microphones into account.
In the proposed method, we use a graph representation of multichannel observations and microphone connections, and then apply the graph Fourier transform (GFT) \cite{Shuman_SPM2013_01}, which enables us to take into account the pairs of microphones that are connected.
%
\begin{figure}[t]
\centering
\includegraphics[width=1.0\columnwidth]{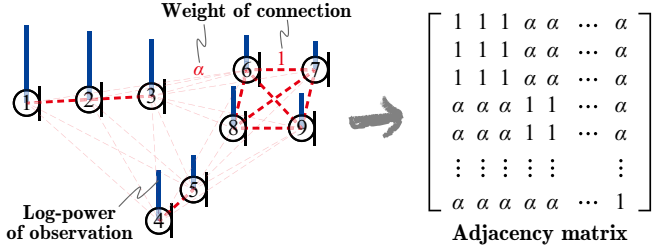}
\caption{Example of observations on undirected graph and relationship between microphone connections and adjacency matrix}
\label{fig:adjacency01}
\end{figure}
%

%
\begin{figure}[t!]
\centering
\includegraphics[width=0.83\columnwidth]{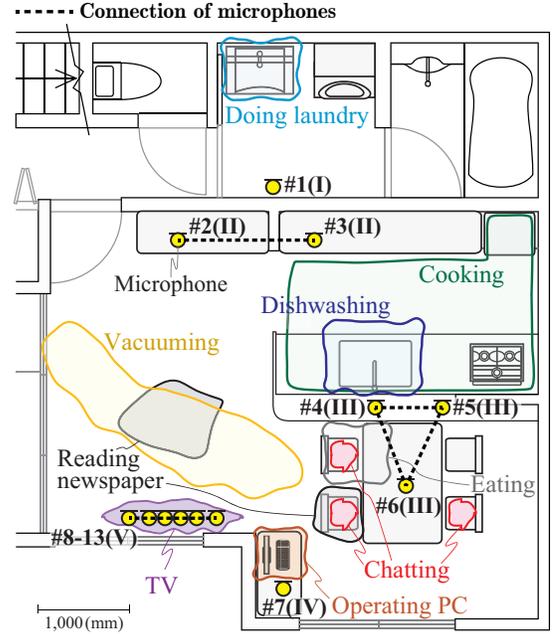}
\caption{Microphone arrangement and sound source locations.
Channel indices (1--13) and group indices of synchronized microphones (I--V) are also indicated. }
\label{fig:condition11}
\end{figure}
\begin{table}[t!]
\vspace{-3pt}
\caption{Conditions of recording and scene classification experiment}
\label{tab:Condition}
\small
\centering
\renewcommand{\arraystretch}{1.0}
\begin{tabular}{lll}
\wcline{1-3}\\[-15pt]
\multicolumn{2}{l}{}&\vspace{-5pt} \\
\multicolumn{2}{l}{Sampling rate}&48 kHz\\[-1pt]
\multicolumn{2}{l}{Quantization bit rate}&16 bits\\[-1pt]
\multicolumn{2}{l}{Sound clip length}&8 s\\[-1pt]
\multicolumn{2}{l}{Frame length\hspace{1pt}/\hspace{1pt}FFT point}&20 ms\hspace{1pt}/\hspace{1pt}2,048\\[-1pt]
\cline{1-3}\\[-10pt]
\multicolumn{2}{l}{Connection weight $\alpha$}&0.01\\[-1pt]
\cline{1-3}\\[-10pt]
\multicolumn{2}{l}{Structure of CNN}&3 conv. \& 3 dense layers\\[-1pt]
\multicolumn{2}{l}{Pooling in CNN layers}&2 $\times$ 2 max pooling\\[-1pt]
\multicolumn{2}{l}{Activation function}&ReLU, softmax (output layer)\\[-1pt]
\multicolumn{2}{l}{\# channels of CNN}&32, 24, 16\\[-1pt]
\multicolumn{2}{l}{\# units of dense layers}&128, 64, 32\\[-1pt]
\multicolumn{2}{l}{Optimizer}&Adam\\[-1pt]
\multicolumn{2}{l}{\# acoustic topics in sATM}&20\\[-1pt]
\wcline{1-3}
\end{tabular}
\vspace{2pt}
\end{table}
%

%
\begin{figure}[t!]
\centering
\includegraphics[width=0.82\columnwidth]{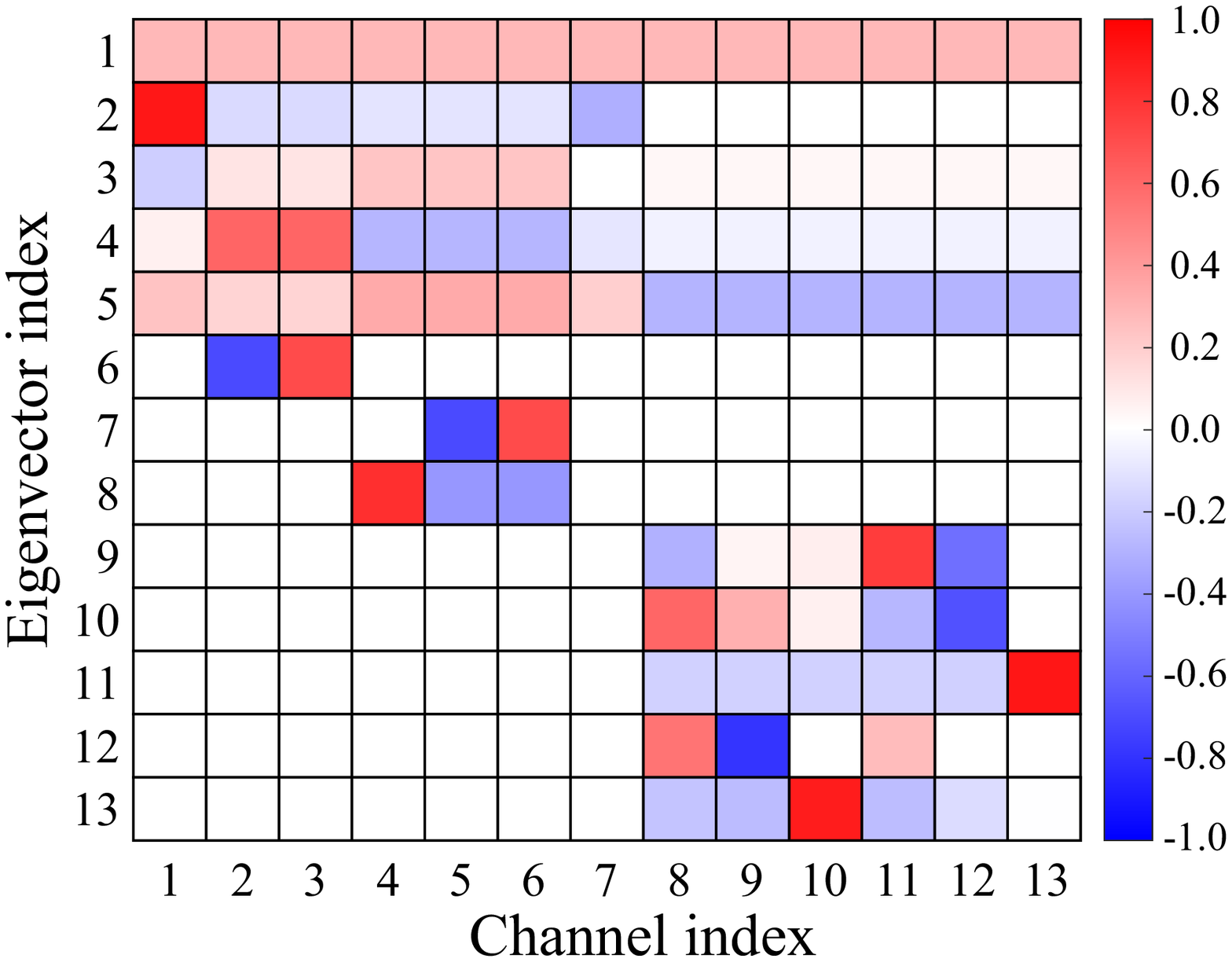}
\vspace{-5pt}
\caption{IGFT matrix ${\bf U}$ in red-blue map representation}
\label{fig:eigenvector}
\vspace{0pt}
\end{figure}
%
%
\begin{figure}[t!]
\centering
\hspace{-5pt}
\subfigure[Weights of 1st-order GC]{%
\includegraphics[width=0.48\columnwidth]{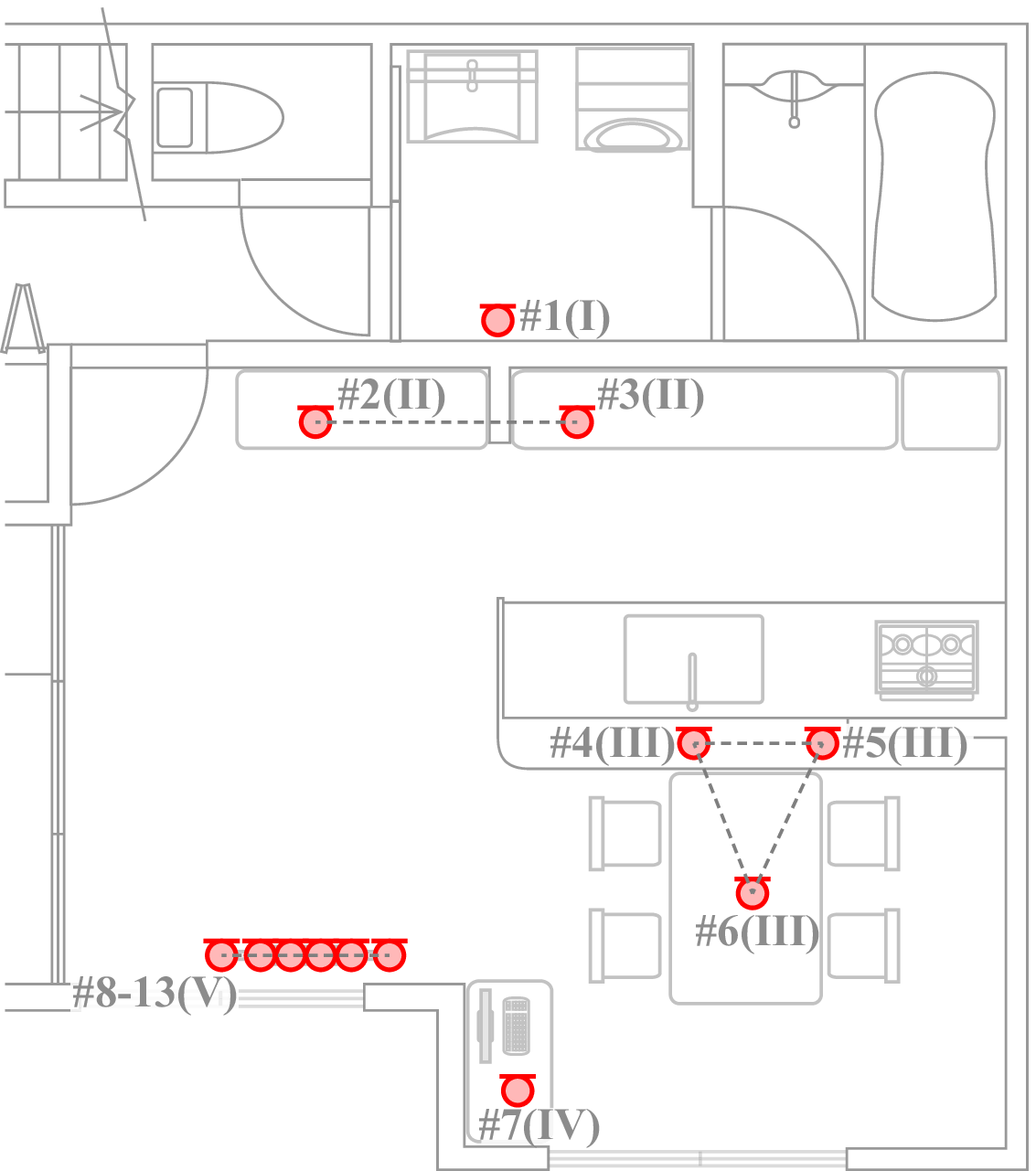}}%
\hspace{7pt}
%
\subfigure[Weights of 5th-order GC]{%
\includegraphics[width=0.48\columnwidth]{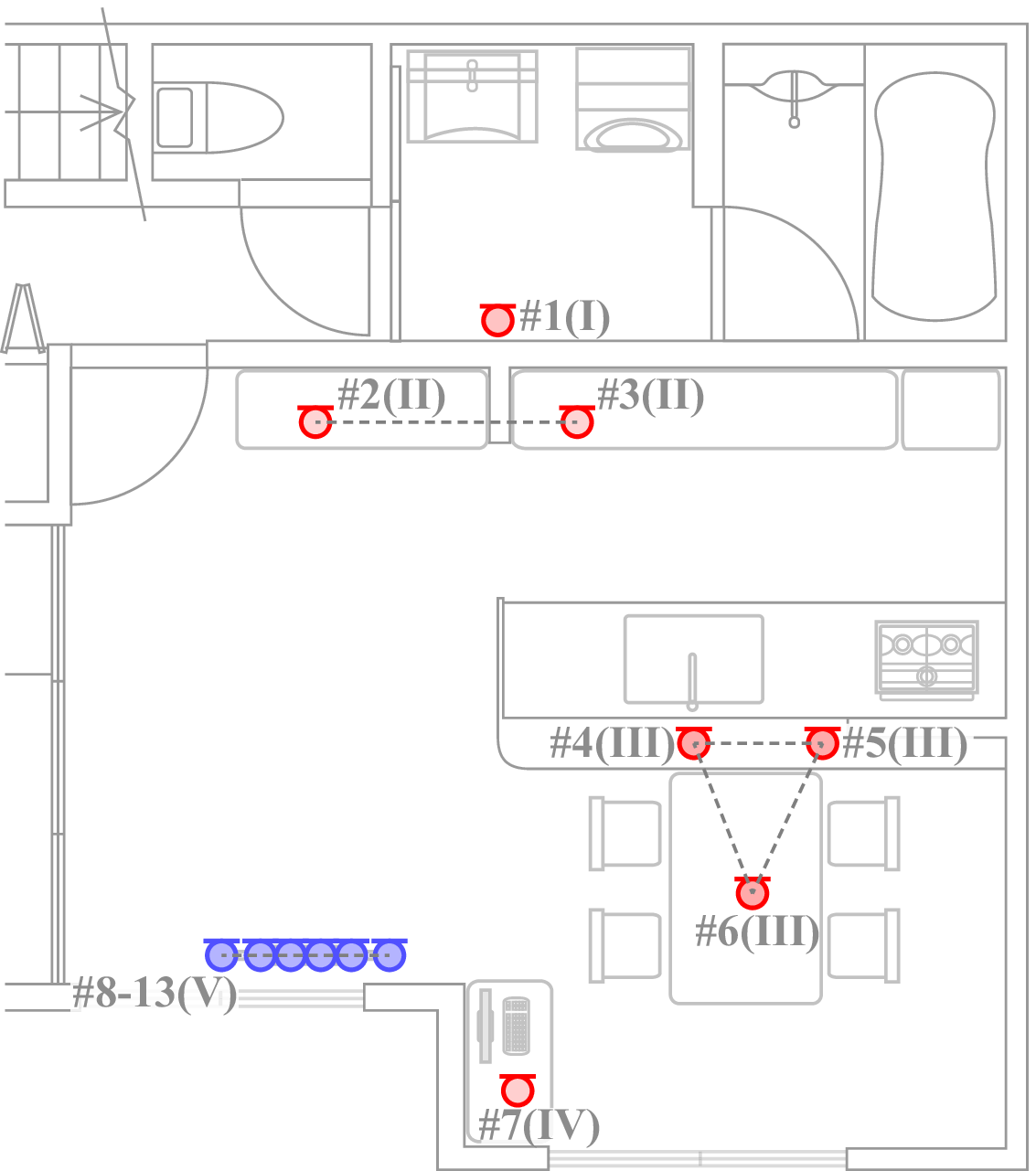}}%
\hspace{0pt}
\subfigure[Weights of 11th-order GC]{%
\includegraphics[width=0.48\columnwidth]{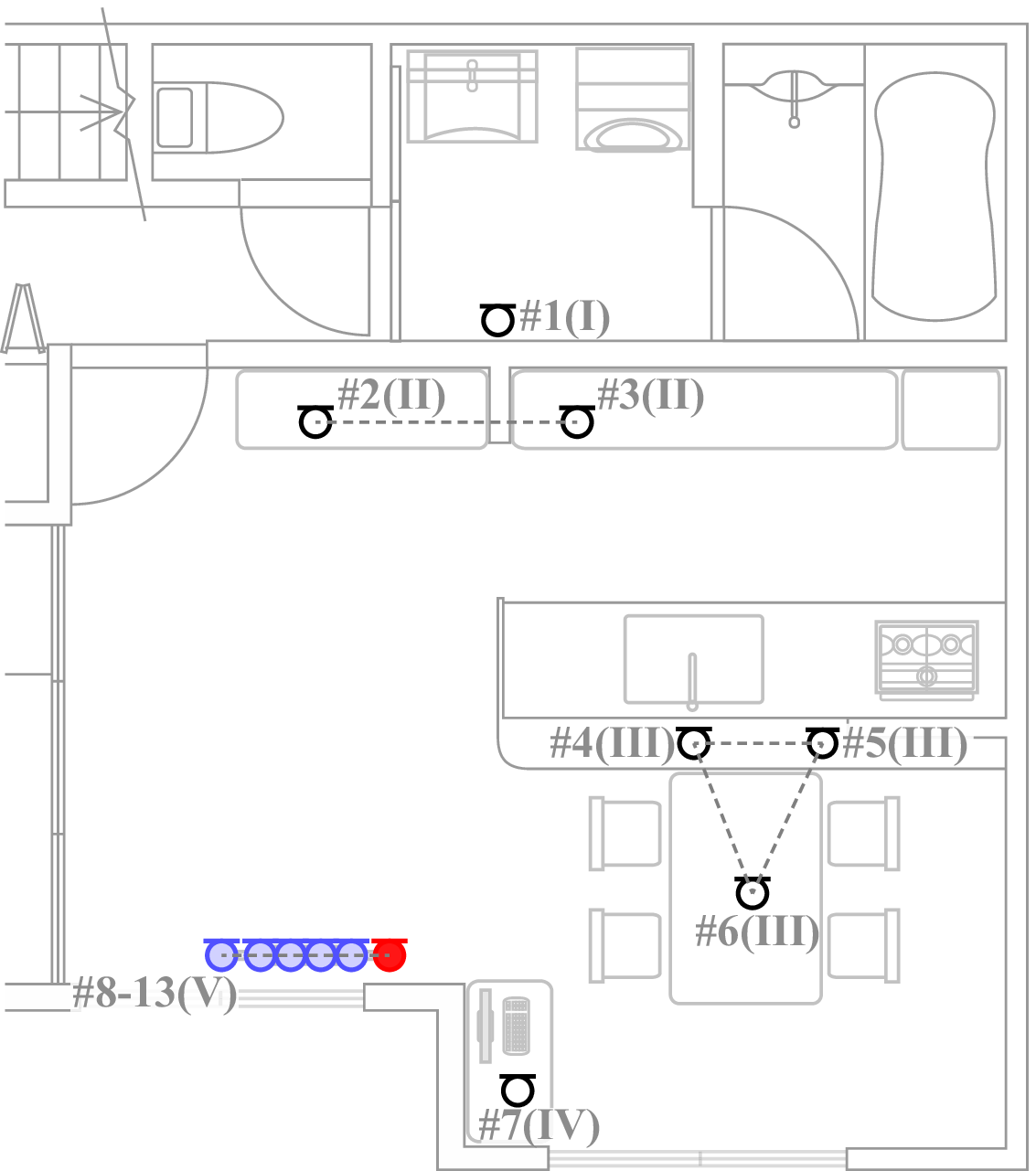}}%
\vspace{0pt}
\caption{Red-blue map representation of {\bf GC} weight $u_{k,n}$ at each microphone}
\label{fig:GCweights}
\end{figure}

We suppose the logarithm powers of the observations on the undirected graph shown in Fig.~\ref{fig:adjacency01}, where the log-power of observation $\log \tilde{a}_{\tau,n}$ and the weight of the microphone connection are represented by the weights of the node and edge, respectively. 
Here, the $N \times N$ adjacency matrix, which represents the weights of the connections between all microphone pairs, is defined as
%
\begin{align}
\hspace{-19pt} A(m,n) = \begin{cases}
    1 & {\rm if\ channels}\ m\ {\rm and}\ n\ {\rm are\ connected}\\
    0 \ {\rm or} \ \alpha & \textit{otherwise}, \end{cases} \!\!\!
\end{align}
%
where $\alpha$ is an arbitrary weight of the connection, which is in the range of 0.0--1.0.
We also consider the $N \times N$ degree matrix {\bf D}, which is a diagonal matrix whose diagonal elements are represented as
%
\begin{align}
{\bf D}(m,m) &= \sum_{n} {\bf A}(m,n).
\end{align}
%
\noindent The degree matrix indicates the number of microphones connected with microphone $m$.
Then, the unweighted graph Laplacian is represented as
%
\begin{align}
{\bf L} &\triangleq {\bf D} - {\bf A},
\end{align}
%
\noindent where {\bf L} is also a symmetric matrix since both {\bf D} and {\bf A} are symmetric matrices.
Thus, the eigendecomposition of ${\bf L}$ is expressed as
%
\begin{align}
{\bf L} &= {\bf U} {\bf \Lambda} {\bf U}^{\mathsf{T}},
\end{align}
%
\noindent where ${\bf U}$ and ${\bf \Lambda}$ are the eigenvector matrix and the diagonal matrix whose diagonal elements $\lambda_{m}$ are equal to the eigenvalues in ascending order, respectively.
The eigenvector matrix ${\bf U}^{\mathsf{T}}$ and its transpose ${\bf U}$ are the GFT and inverse graph Fourier transform (IGFT) matrices, respectively, which enable the basis transformations considering the connections between microphones.

%
\begin{figure}[t!]
\centering
\includegraphics[width=0.82\columnwidth]{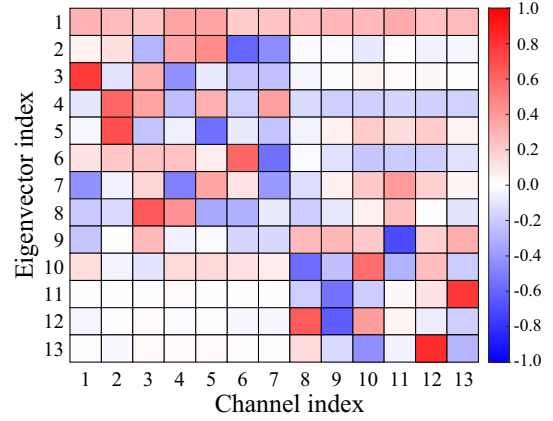}
\vspace{-5pt}
\caption{{\bf SC} matrix in red-blue map representation}
\label{fig:SCmatrix}
\vspace{0pt}
\end{figure}
%
%
\begin{figure}[t!]
\centering
\hspace{-5pt}
\subfigure[Weights of 1st-order SC]{%
\includegraphics[width=0.48\columnwidth]{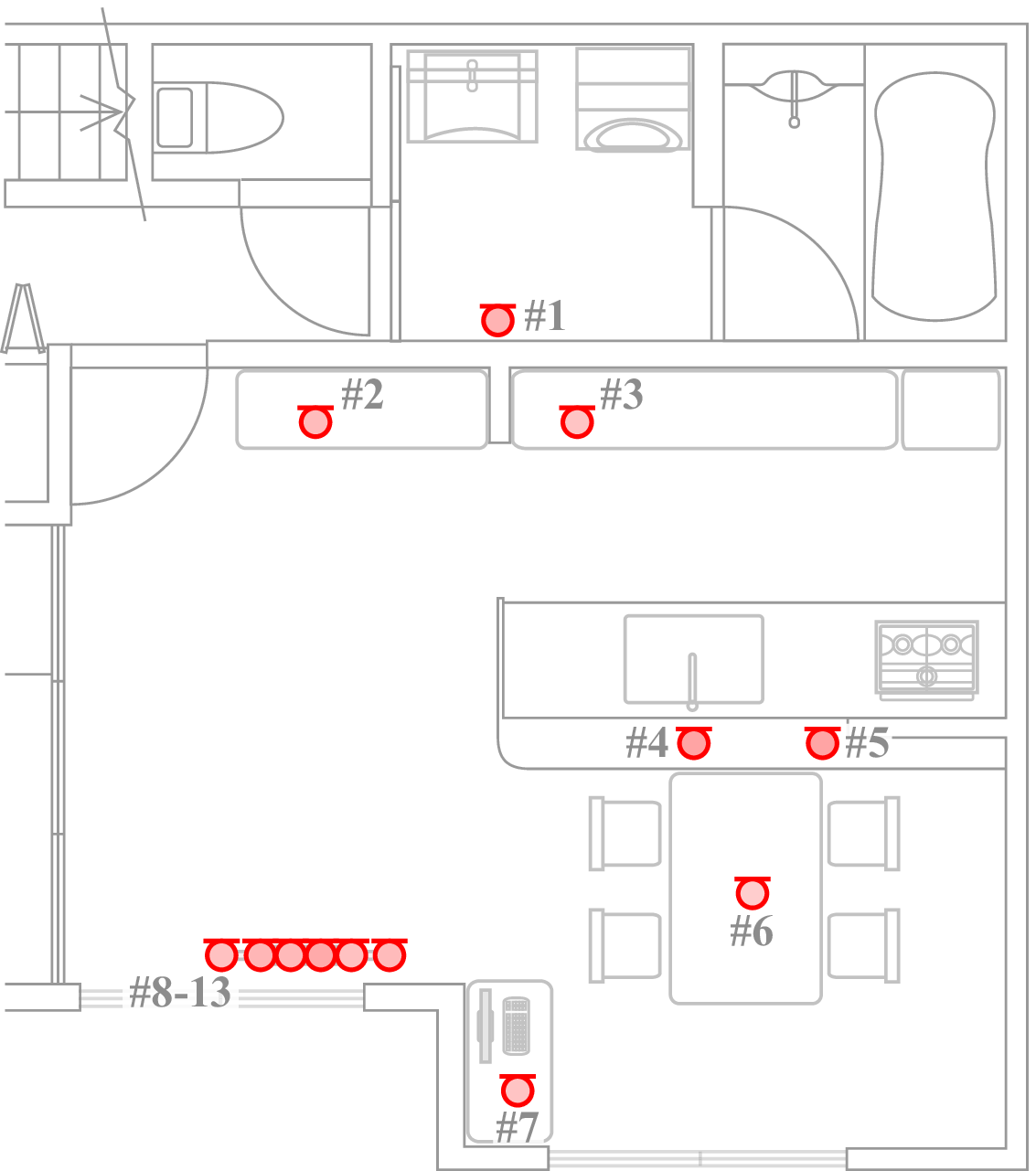}}%
\hspace{7pt}
%
\subfigure[Weights of 5th-order SC]{%
\includegraphics[width=0.48\columnwidth]{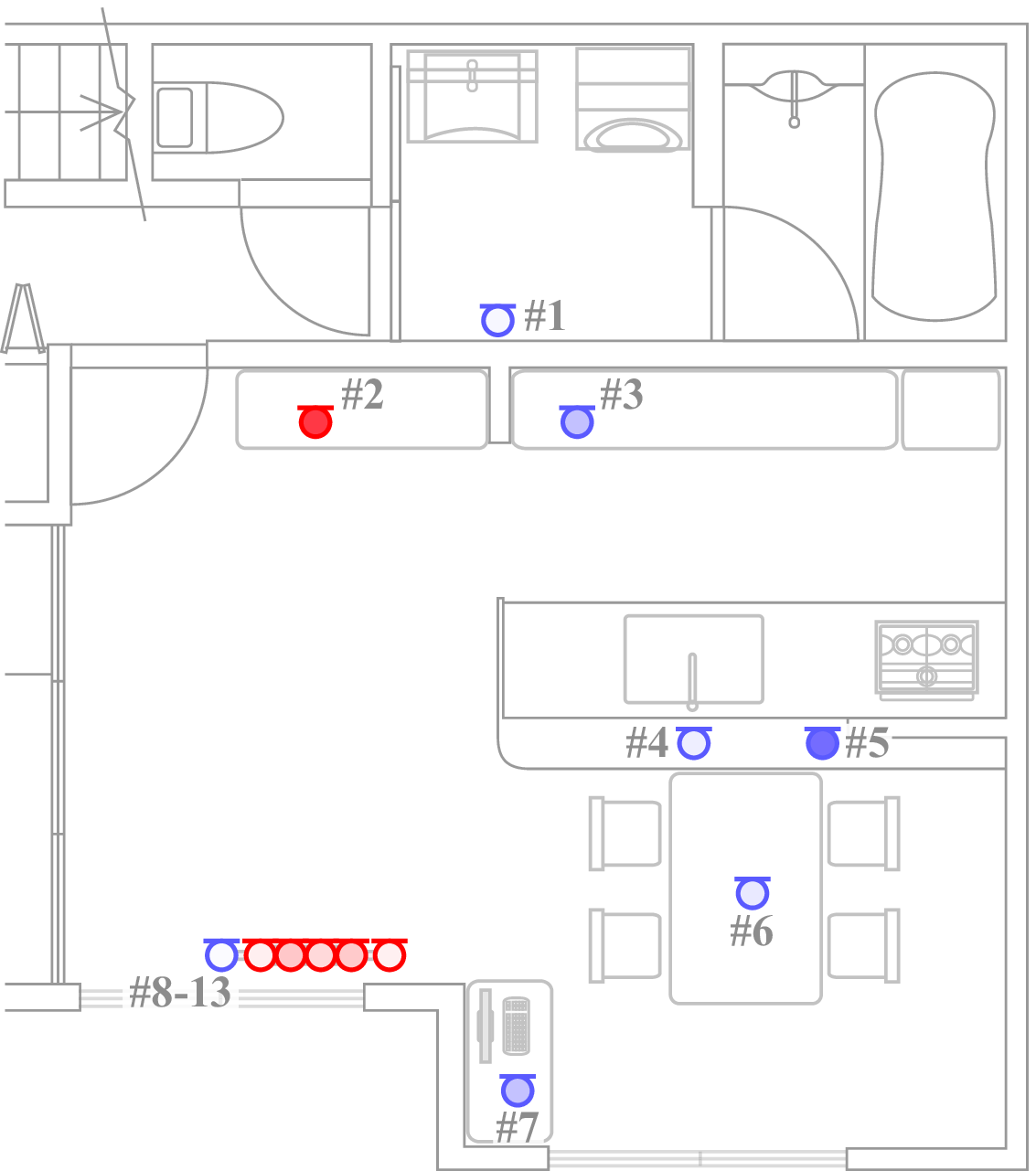}}%
\hspace{0pt}
\subfigure[Weights of 11th-order SC]{%
\includegraphics[width=0.48\columnwidth]{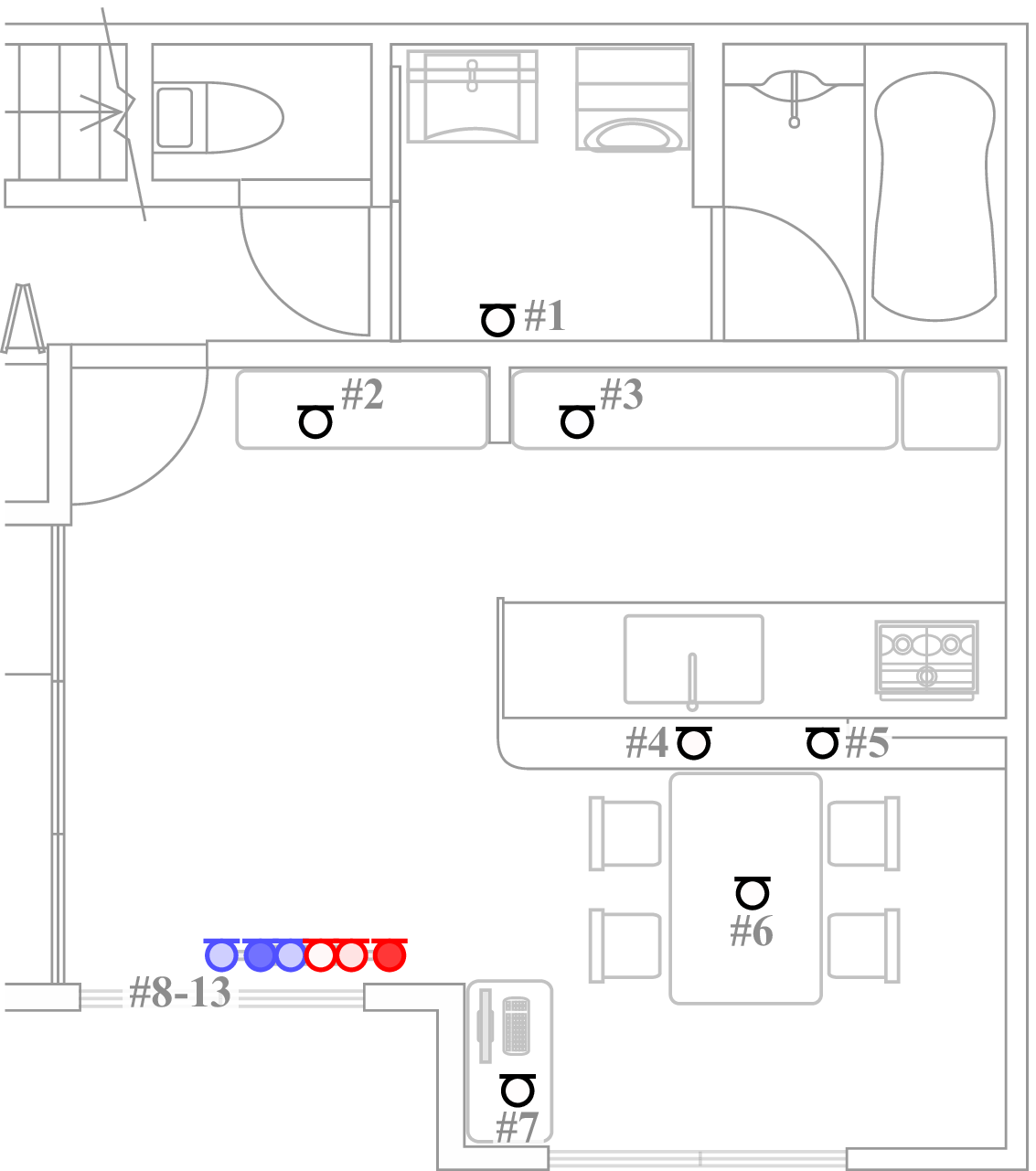}}%
\vspace{-3pt}
\caption{Red-blue map representation of {\bf SC} weight $e_{k,n}$ at each microphone}
\label{fig:SCweights}
\end{figure}

Thus, the proposed spatial feature, in which the connections between microphones are considered, is defined in terms of the IGFT of the log-amplitude vector ${\bf q}_{\tau}$ as

%
\begin{align}
{\bf e}_{\tau}={\bf U} {\bf q}_{\tau}.
  \label{eq:def_gSC}
\end{align}
%

Because the proposed spatial feature also resembles the conventional cepstrum and SC as described in Appendix, we call it the graph cepstrum (GC).
The GC can extract a spatial pattern without any training data, which are required for the extraction of the SC if we have information on microphone connections.
%
%
%
\section{Evaluation Experiments}
\label{sec:Experiments}
\subsection{Experimental Conditions}
\label{ssec:cond}
To evaluate characteristics of the GC and its effectiveness for acoustic scene analysis, we investigated the behavior of the GC and conducted classification experiments on acoustic scenes using spatial information.
Since most of the public datasets for acoustic scene analysis including TUT Acoustic Scenes 2017 \cite{Mesaros_DCASE2017_01} and AudioSet \cite{Gemmeke_ICASSP2017_01} are provided in single or stereo channels, we recorded a multichannel sound dataset with 13 synchronized microphones in a real environment.
The sound dataset includes nine acoustic scenes, ``vacuuming,'' ``cooking,'' ``dishwashing,'' ``eating,'' ``reading a newspaper,'' ``operating a PC,'' ``chatting,'' ``watching TV,'' and ``doing the laundry,'' which occur frequently in a living room.
The sound source locations and the microphone arrangement are shown in Fig.~\ref{fig:condition11}.
Each arabic number and roman number indicate the microphone index and the group index of synchronized microphones, respectively.
The sound dataset consists of 257.1 min recordings, which were randomly separated into 5,180 sound clips for model training and 2,532 sound clips for classification evaluation, where no acoustic scene overlapped with another scene in all the sound clips.
To evaluate the scene classification performance using microphones with synchronization mismatch among the microphone groups, the recorded sounds for classification evaluation were misaligned with various error times among the microphone groups shown in Fig.~\ref{fig:condition11}.
The error times were randomly sampled from a Gaussian distribution with $\mu = 0$ and various variances $\sigma^{2}$.
The other recording and experimental conditions are listed in Table~\ref{tab:Condition}.
%
%
%
%
\begin{figure*}[t!]
\centering
\hspace{-5pt}
\subfigure[IGFT matrix for $\alpha = 0.1$]{%
\includegraphics[width=0.666\columnwidth]{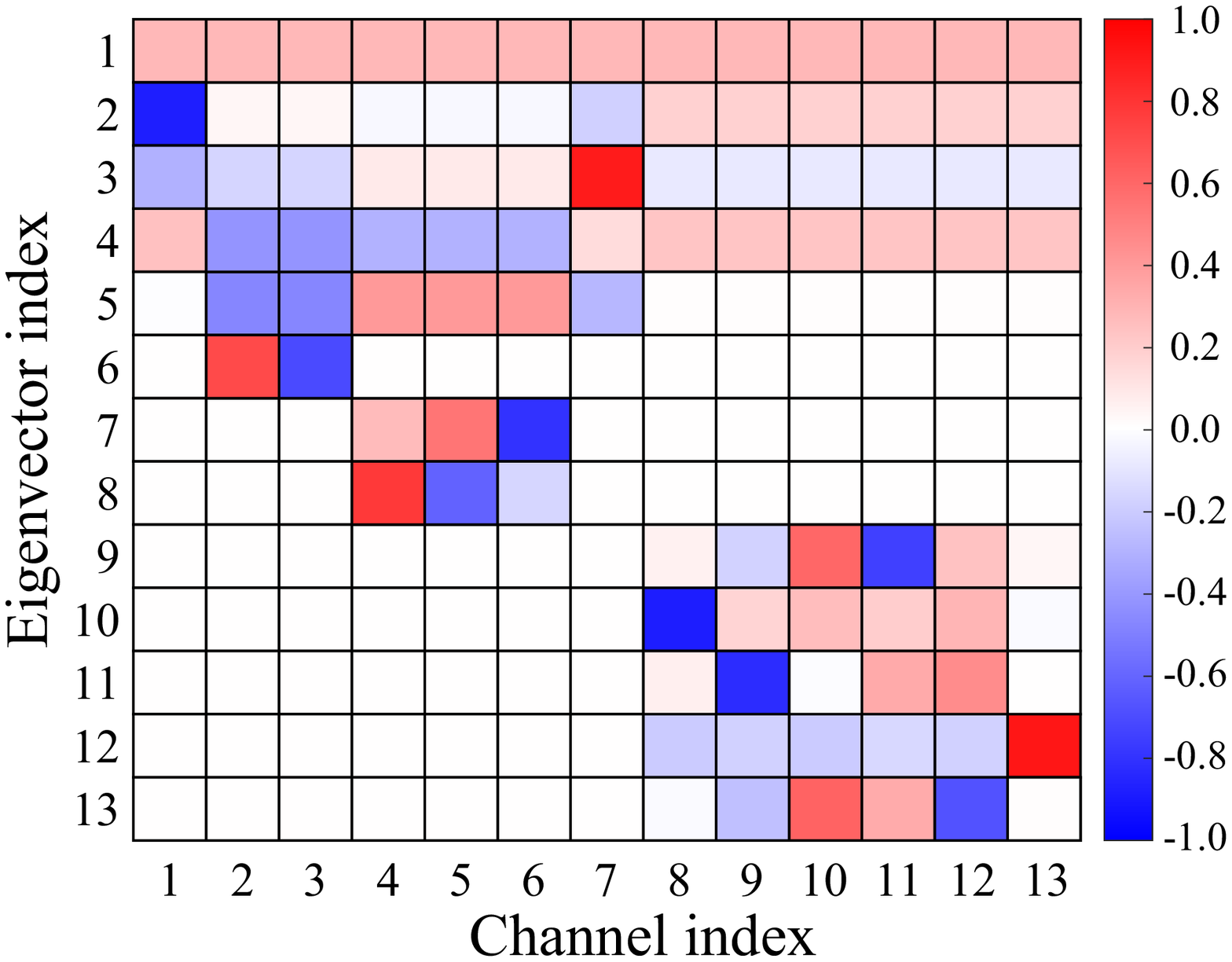}}%
\hspace{7pt}
%
\subfigure[IGFT matrix for $\alpha = 0.001$]{%
\includegraphics[width=0.666\columnwidth]{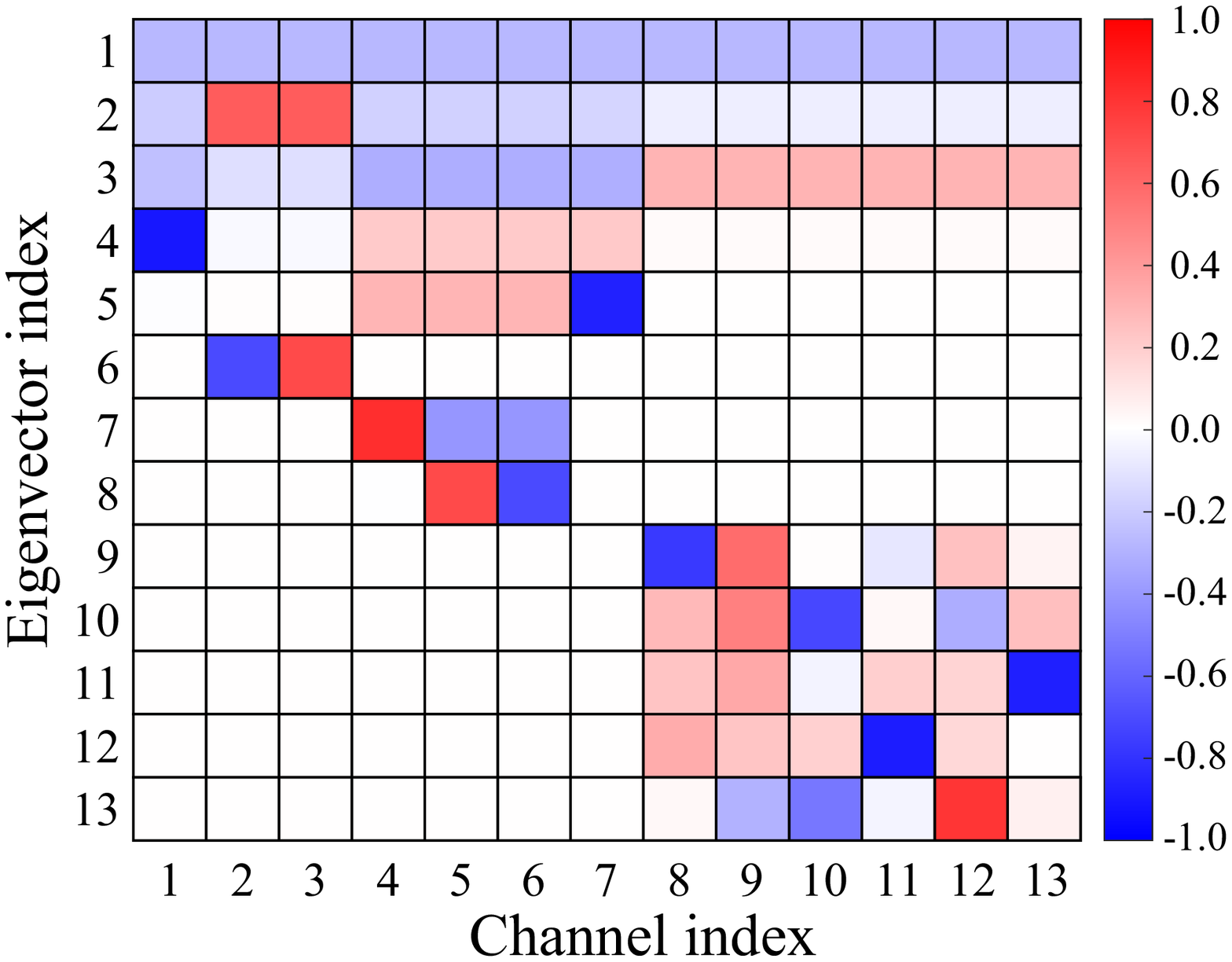}}%
\hspace{7pt}
\subfigure[IGFT matrix for $\alpha = 0.0001$]{%
\includegraphics[width=0.666\columnwidth]{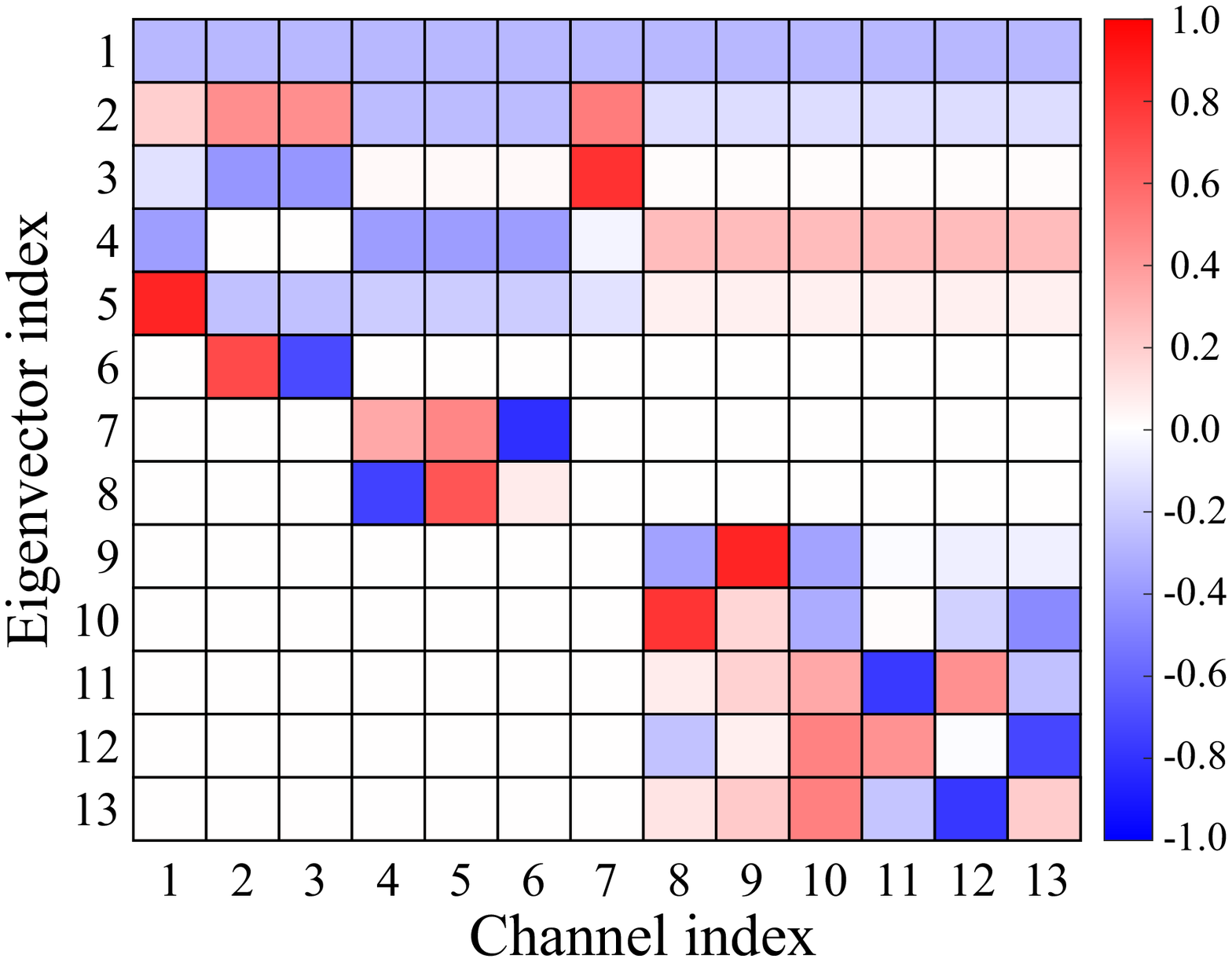}}%
\vspace{-6pt}
\caption{IGFT matrix ${\bf U}$ in red-blue map representation with various connection weights $\alpha$}
\label{fig:GCweightMatrices}
\vspace{-2pt}
\end{figure*}
%
%
\subsection{Spatial Information Extracted from Graph Cepstrum}
\label{ssec:spatialinfo}
To investigate how the GC extracts spatial information, we examined the IGFT matrix ${\bf U}$ calculated using the connection weight $A(m,n)$, which is 1.0 when microphones are connected, and $0.01$ otherwise.
This is because sounds can propagate to all of the microphones, and it is expected that the observations of microphones will be slightly correlated.

Fig.~\ref{fig:eigenvector} shows the IGFT matrix ${\bf U}$ as a red-blue map.
The $k$th-row vector of ${\bf U}$ corresponds to the $k$th eigenvector of the graph Laplacian ${\bf L}$.
The $k$th-order GC is calculated using the $k$th-row vector of ${\bf U}$ as follows:

%
\begin{align}
e_{\tau, k} = {\bf u}_{k} {\bf q}_{\tau} = \sum_{n=1}^{N} u_{k, n} q_{\tau, n},
\label{eq:etaun}
\end{align}
%

\noindent where $e_{\tau, k}$, ${\bf u}_{k}$, $u_{k, n}$, and $q_{\tau, n}$ are the $k$th-order GC, the $k$th-row vector of ${\bf U}$, the $(k, n)$ entry of ${\bf U}$, and the $n$th element of ${\bf q}_{\tau}$, respectively.
This indicates that the $k$th-order GC is obtained by a linear combination of log-amplitudes $q_{\tau, n}$, where $u_{k, n}$ is the weight of the linear combination.
In Fig.~\ref{fig:GCweights}, we also show the weights at each microphone position to investigate how the GC extracts spatial information.

From Figs.~\ref{fig:eigenvector} and \ref{fig:GCweights}, it can be interpreted that the first-order GC represents the average sound level in the whole space because all the weights ${\bf u}_{k}$ are positive and the same values.
For the middle-order GC (from second order to fifth order), the signs of the weights among the closely located microphones are similar.
This means that the middle-order GC can extract rough spatial information as the middle-order cepstrum coefficients capture the spectral envelope.
%
%
For the higher-order GC, the weights of only part of the connected microphone group are active and the summation of the weights $u_{k,n}$ in the microphone groups is close to zero.
When a sound source is far from the microphone group, the log amplitudes $q_{\tau,n}$ are approximately equal between the microphones, and $e_{\tau, k} $ becomes close to zero.
Thus, the higher-order GC coefficients capture spatial information of sound sources close to the microphone group.
\begin{table}[t!]
\caption{Similarity between IGFT and SC matrices}
\label{tab:Similarity}
\footnotesize
\centering
\renewcommand{\arraystretch}{1.0}
\begin{tabular}{clrrrrrr}
\wcline{1-8}\\[-8pt]
\!\!&&\multicolumn{5}{c}{{\bf IGFT}}&\multicolumn{1}{c}{\multirow{2}{*}{\!{\bf SC}}}\!\!\\
\cline{3-7}\\[-8pt]
\!\!&\!\!\!&\!\!$\alpha=1.0$\!\!\!&\multicolumn{1}{c}{\!\!\!0.1\!\!\!}&\multicolumn{1}{c}{\!\!\!\!0.01\!\!\!}&\multicolumn{1}{c}{\!\!\!\!0.001\!\!\!}&\multicolumn{1}{c}{\!\!\!\!0.00001\!\!\!}&\!\!\!\!\\
\wcline{1-8}\\[-7pt]
\!\!\multirow{6}{*}{{\bf IGFT}}&\!\!\!\!$\alpha\!=\!1.0$\!\!\!\!\!&\!\!\!\!\! - \!\!\!\!&\!\!\!\! 16.802 \!\!\!\!&\!\!\!\! 15.469 \!\!\!\!&\!\!\!\! 16.616 \!\!\!\!&\!\!\!\! 17.005 \!\!\!\!&\!\!\! 10.741\!\!\\[1pt]
\!\!&\!\!\!\!\!$\alpha\!=\!0.1$\!\!\!\!            &\!\!\!\!\!16.802 \!\!\!\!&\!\!\!\! -    \!\!\!\!&\!\!\!\! 5.584 \!\!\!\!&\!\!\!\! 10.209 \!\!\!\!&\!\!\!\! 5.252 \!\!\!\!&\!\!\! 9.385 \!\!\!\\[1pt]
\!\!&\!\!\!\!\!$\alpha\!=\!0.01$\!\!\!\!          &\!\!\!\!\!15.469 \!\!\!\!&\!\!\!\! 5.584 \!\!\!\!&\!\!\!\! - \!\!\!\!&\!\!\!\! 9.450 \!\!\!\!&\!\!\!\! 8.695 \!\!\!\!&\!\!\! 8.926 \!\!\!\\[1pt]
\!\!&\!\!\!\!\!$\alpha\!=\!0.001$\!\!\!\!        &\!\!\!\!\!16.616 \!\!\!\!&\!\!\!\! 10.209 \!\!\!\!&\!\!\!\! 9.450 \!\!\!\!&\!\!\!\! - \!\!\!\!&\!\!\!\! 7.875 \!\!\!\!&\!\!\! 10.044 \!\!\!\\[1pt]
\!\!&\!\!\!\!\!$\alpha\!=\!0.0001$\!\!\!\!      &\!\!\!\!\!17.005 \!\!\!\!&\!\!\!\! 5.252 \!\!\!\!&\!\!\!\! 8.695 \!\!\!\!&\!\!\!\! 7.875 \!\!\!\!&\!\!\!\! - \!\!\!\!&\!\!\! 10.359 \!\!\!\\[2pt]
\multicolumn{2}{c}{{\bf SC}}\!\!\!\!&\!\!\!\!\!10.741 \!\!\!\!&\!\!\!\! 9.385 \!\!\!\!&\!\!\!\! 8.926 \!\!\!\!&\!\!\!\! 10.044 \!\!\!\!&\!\!\!\! 10.359 \!\!\!\!&\!\!\! - \!\!\\
\wcline{1-8}
\end{tabular}
\end{table}
%
%
\subsection{Comparison of Spatial Information Extracted from Graph Cepstrum and Spatial Cepstrum}
\label{ssec:comparison}
To compare spatial information extracted from the GC with the SC, we also show the transformation matrix ${\bf E}^{\mathsf{T}}$ of SC (referred to as a SC matrix) as a red-blue map in Figs.~\ref{fig:SCmatrix} and \ref{fig:SCweights}.
To calculate the SC matrix, the synchronized original recordings were used.
The $k$th-row vector of the SC matrix corresponds to the $k$th eigenvector of the covariance matrix ${\bf R}_{q}$ of ${\bf q}_{\tau}$.
The $k$th-order SC is calculated using the $k$th-row vector of ${\bf E}^{{\mathsf T}}$ as follows:

%
\begin{align}
d_{\tau, k} &= {\bf e}_{k}^{\mathsf{T}} {\bf q}_{\tau} = \sum_{n=1}^{N} e_{k, n} q_{\tau, n},
\end{align}

\noindent where $d_{\tau, k}$, ${\bf e}_{k}^{\mathsf{T}}$, $e_{k, n}$, and $q_{\tau, n}$ are the $k$th-order SC, the $k$th-row vector of ${\bf E}^{\mathsf{T}}$, the $(k, n)$ entry of ${\bf E}^{\mathsf{T}}$, and the $n$th element of ${\bf q}_{\tau}$, respectively.
Similarly to the GC, the $k$th-order SC is calculated as a linear combination of log-amplitudes, in which $e_{k,n}$ act as the weights of the linear combination.

Considering that the signs of the eigenvalue decomposition can be flipped, Figs.~\ref{fig:eigenvector}--\ref{fig:SCweights} show that the GC and SC extract similar spatial information.
Thus, the experimental results show that the GC can extract a similar spatial pattern to the SC without any training data for learning the spatial pattern, which is required for the extraction of the SC.
%
%
%
\vspace{-5pt}
\subsection{Spatial Information Extracted from Graph Cepstrum with Various Connection Weights}
\label{ssec:GCmatrices}
\vspace{-5pt}
Further investigation of spatial information extracted from GC with various connection weights $\alpha$ was conducted.
In this experiment, we calculated the IGFT matrix ${\bf U}$ using the connection weight $A(m,n)$, which is 1.0 when microphones are connected and $\alpha = 0.1, 0.001$, and $0.0001$ otherwise.
Figs.~\ref{fig:GCweightMatrices}~(a)--\ref{fig:GCweightMatrices}~(c) show the IGFT matrix ${\bf U}$ with $\alpha = 0.1, 0.001$, and $0.0001$, respectively.
Considering that the signs of the eigenvalue decomposition can be flipped, Figs.~\ref{fig:eigenvector} and \ref{fig:GCweightMatrices} indicate that the GC extracts spatial patterns without strongly depending on $\alpha$ because the active channels are similar for IGFT matrices with various $\alpha$.

We then evaluated the similarity between IGFT and SC matrices in terms of the sum of squares error calculated as follows:

\vspace{-6pt}
\begin{align}
r_{\alpha_{1} \alpha_{2}} &= \sum_{i,j=1}^{N}  (|u_{i, j}^{(\alpha_{1})}| - |u_{i, j}^{(\alpha_{2})}|)^{2},\\
r_{\alpha_{1} SC} &= \sum_{i,j=1}^{N}  (|u_{i, j}^{(\alpha_{1})}| - |e_{i, j}|)^{2},
\end{align}

\noindent where $r_{\alpha_{1} \alpha_{2}}$ is the sum of squares error between IGFT matrices whose $\alpha$ values are $\alpha_{1}$ and $\alpha_{2}$.
$r_{\alpha_{1} SC}$ indicates the sum of squares error between IGFT and SC matrices.
For comparison, we evaluated the similarity using $\alpha = 1.0$, which indicates that all microphones are connected.
As shown in Table~\ref{tab:Similarity}, when $\alpha_{1}$ and $\alpha_{2}$ are $0.1, 0.01, 0.001$, and $0.0001$, the IGFT matrices are more similar to each other than the SC matrix and the IGFT matrix with $\alpha = 1.0$.
Thus, the GC can extract spatial patterns with robustness against $\alpha$.
%
%
\subsection{Acoustic Scene Classification Utilizing Spatial Information}
\label{ssec:classification}
To evaluate the effectiveness of the GC for acoustic scene analysis, scene classification experiments were conducted.
As the acoustic feature, the 13-dimensional GC coefficients and 512-dimensional bag-of-acoustic words (BoW) \cite{Imoto_IEICE2016_01,Imoto_EUSIPCO2017_01} calculated from the GC were used.
Acoustic scenes were then modeled and classified with respect to each sound clip using a GMM, a supervised acoustic topic model (sATM) \cite{Imoto_IEICE2016_01,Imoto_EUSIPCO2017_01}, and a CNN.
Specifically, the GMM was applied to acoustic feature vectors ${\bf e}_{\tau}$ and ${\bf d}_{\tau}$ for each acoustic scene $x$, and acoustic scene $x$ of sound clip $c$ was estimated by calculating the product of the likelihoods over the sound clip as follows:
%
\begin{align}
x_{c}=\argmax_{x} \prod^{T_{c}}_{\tau=1} p_{\tau}({\bf f}_{\tau}|x),
  \label{eq:argmax}
\end{align}
%

\noindent where $T_{c}$, ${\bf f}_{\tau}$, and $p_{\tau}({\bf f}_{\tau}|x)$ are the number of frames in sound clip $c$, an acoustic feature vector at time frame $\tau$ such as ${\bf d}_{\tau}$ or ${\bf e}_{\tau}$, and the likelihood of acoustic scene $x$, respectively.
The sATM is a method for ASA based on a Bayesian generative model, which learns the relationship between acoustic scenes and the event sequence.
As other methods of acoustic scene classification utilizing a distributed microphone array, we also evaluated classifiers using a classifier stacking-based method \cite{Kurby_DCASE2016_01}.
These acoustic features and classifiers were selected with reference to \cite{Imoto_TASLP2017_01}.

\begin{figure}[t!]
\centering
\includegraphics[width=1.0\columnwidth]{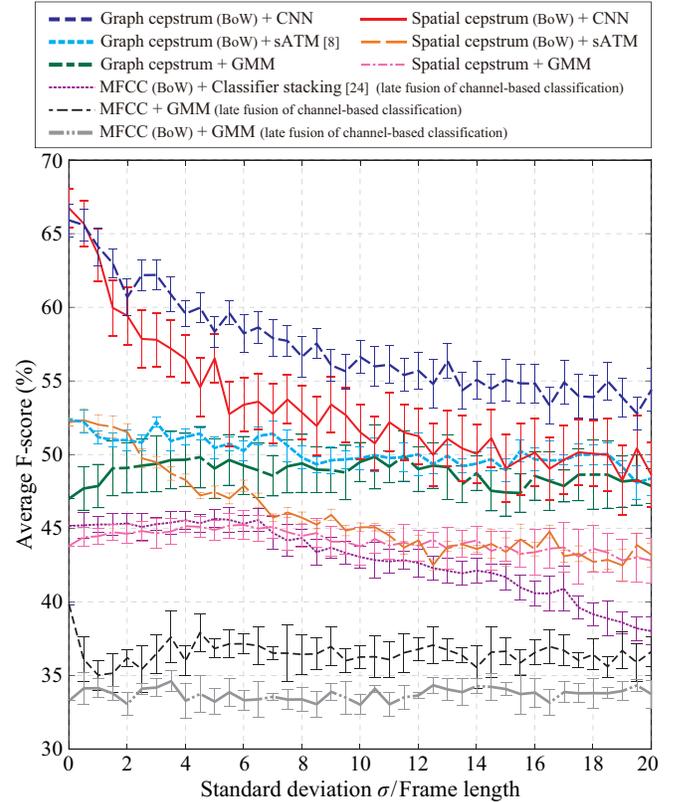}
\caption{Acoustic scene classification accuracy with various synchronization error times between connected microphone groups}
\label{fig:gcres01}
\end{figure}

The performance of classifying acoustic scenes is shown in Fig.~\ref{fig:gcres01}.
For each experimental condition, the acoustic scene modeling and classification were conducted ten times with various synchronization error times sampled randomly.
The results show that when the synchronization error between microphone groups is small, the GC and conventional SC effectively classify acoustic scenes.
When the synchronization error between microphone groups increases, the scene classification performance of the GC slightly decreases.
This is because the proposed GC less considers weakly connected microphones; therefore, the GC is less affected by the synchronization error and is more robust against synchronization error.
In contrast, the classification accuracy decreases rapidly when using the conventional SC.
This is because only the sounds recorded by the synchronized microphones were used to obtain the eigenvector matrix ${\bf E}$; thus, the synchronization error led to an inappropriate basis transformation.
%
%
%
%
\section{Conclusion}
\label{sec:conclude}
In this paper, an effective spatial feature extraction method using partially synchronized and/or closely located distributed microphones was proposed.
In the proposed method, the graph cepstrum (GC), which is defined as the inverse graph Fourier transform of the logarithm power of a multichannel observation, is derived.
Then it was demonstrated that the GC in a ring graph is identical to the cepstrum and spatial cepstrum in a circularly symmetric microphone arrangement with an isotropic sound field.
The experimental results show that the proposed GC robustly extracts spatial information with consideration of the microphone connections.
Moreover, the experiments using real environmental sounds showed that the GC enables more robust classification of acoustic scenes than conventional spatial features even when the synchronization mismatch between partially synchronized microphone groups is large.
%
%
%
\small
\bibliographystyle{IEEEbib}
\bibliography{IEEEabrv,IEICE2019ref,KeisukeImoto09}

\begin{thebibliography}{10}

\bibitem{Imoto_AST2018_01}
K.~Imoto,
\newblock ``Introduction to acoustic event and scene analysis,''
\newblock {\em Acoustical Science and Technology}, vol. 39, no. 3, pp.
  182--188, 2018.

\bibitem{Peng_ICME2009_01}
Y.~Peng, C.~Lin, M.~Sun, and K.~Tsai,
\newblock ``Healthcare audio event classification using hidden {M}arkov models
  and hierarchical hidden {M}arkov models,''
\newblock {\em Proc. {IEEE} International Conference on Multimedia and Expo
  {\rm (}ICME{\rm )}}, pp. 1218--1221, 2009.

\bibitem{Guyot_ICASSP2013_01}
P.~Guyot, J.~Pinquier, and R.~Andr\'{e}-Obrecht,
\newblock ``Water sound recognition based on physical models,''
\newblock {\em Proc. {IEEE} International Conference on Acoustics, Speech and
  Signal Processing {\rm (}ICASSP{\rm )}}, pp. 793--797, 2013.

\bibitem{Harma_ICME2005_01}
A.~Harma, M.~F. McKinney, and J.~Skowronek,
\newblock ``Automatic surveillance of the acoustic activity in our living
  environment,''
\newblock {\em Proc. {IEEE} International Conference on Multimedia and Expo
  {\rm (}ICME{\rm )}}, 2005.

\bibitem{Radhakrishnan_WASPAA2005_01}
R.~Radhakrishnan, A.~Divakaran, and P.~Smaragdis,
\newblock ``Audio analysis for surveillance applications,''
\newblock {\em Proc. 2005 {IEEE} Workshop on Applications of Signal Processing
  to Audio and Acoustics {\rm (}WASPAA{\rm )}}, pp. 158--161, 2005.

\bibitem{Ntalampiras_ICASSP2009_01}
S.~Ntalampiras, I.~Potamitis, and N.~Fakotakis,
\newblock ``On acoustic surveillance of hazardous situations,''
\newblock {\em Proc. {IEEE} International Conference on Acoustics, Speech and
  Signal Processing {\rm (}ICASSP{\rm )}}, pp. 165--168, 2009.

\bibitem{Eronen_TASLP2006_01}
A.~Eronen, V.~T. Peltonen, J.~T. Tuomi, A.~P. Klapuri, S.~Fagerlund, T.~Sorsa,
  G.~Lorho, and J.~Huopaniemi,
\newblock ``Audio-based context recognition,''
\newblock {\em {IEEE} Trans. Audio Speech Lang. Process.}, vol. 14, no. 1, pp.
  321--329, 2006.

\bibitem{Imoto_IEICE2016_01}
K.~Imoto and S.~Shimauchi,
\newblock ``Acoustic scene analysis based on hierarchical generative model of
  acoustic event sequence,''
\newblock {\em IEICE Trans. Inf. Syst.}, vol. E99-D, no. 10, pp. 2539--2549,
  2016.

\bibitem{Schroder_ICASSP2016_01}
J.~Schr{\"o}der, J.~Anemiiller, and S.~Goetze,
\newblock ``Classification of human cough signals using spectro-temporal
  {G}abor filterbank features,''
\newblock {\em Proc. IEEE International Conference on Acoustics, Speech and
  Signal Processing {\rm (}ICASSP{\rm )}}, pp. 6455--6459, 2016.

\bibitem{Zhang_TASLP2001_01}
T.~Zhang and C.~J. Kuo,
\newblock ``Audio content analysis for online audiovisual data segmentation and
  classification,''
\newblock {\em {IEEE} Trans. Audio Speech Lang. Process.}, vol. 9, no. 4, pp.
  441--457, 2001.

\bibitem{Jin_INTERSPEECH2012_01}
Q.~Jin, P.~F. Schulam, S.~Rawat, S.~Burger, D.~Ding, and F.~Metze,
\newblock ``Event-based video retrieval using audio,''
\newblock {\em Proc. INTERSPEECH}, 2012.

\bibitem{Ohishi_ICASSP2013_01}
Y.~Ohishi, D.~Mochihashi, T.~Matsui, M.~Nakano, H.~Kameoka, T.~Izumitani, and
  K.~Kashino,
\newblock ``{B}ayesian semi-supervised audio event transcription based on
  {M}arkov {I}ndian buffet process,''
\newblock {\em Proc. {IEEE} International Conference on Acoustics, Speech and
  Signal Processing {\rm (}ICASSP{\rm )}}, pp. 3163--3167, 2013.

\bibitem{Liang_ICASSP2017_01}
J.~Liang, L.~Jiang, and A.~Hauptmann,
\newblock ``Temporal localization of audio events for conflict monitoring in
  social media,''
\newblock {\em Proc. IEEE International Conference on Acoustics, Speech and
  Signal Processing {\rm (}ICASSP{\rm )}}, pp. 1597--1601, 2017.

\bibitem{Mesaros_EUSIPCO2010_01}
A.~Mesaros, T.~Heittola, A.~Eronen, and T.~Virtanen,
\newblock ``Acoustic event detection in real life recordings,''
\newblock {\em Proc. 18th European Signal Processing Conference {\rm
  (}EUSIPCO{\rm )}}, pp. 1267--1271, 2010.

\bibitem{Han_DCASE2017_01}
Y.~Han, J.~Park, and K.~Lee,
\newblock ``Convolutional neural networks with binaural representations and
  background subtraction for acoustic scene classification,''
\newblock {\em Proc. Detection and Classification of Acoustic Scenes and Events
  Workshop {\rm (}DCASE{\rm )}}, pp. 1--5, 2017.

\bibitem{Jallet_DCASE2017_01}
H.~Jallet, E.~\c{C}ak{\i}r, and T.~Virtanen,
\newblock ``Acoustic scene classification using convolutional recurrent neural
  networks,''
\newblock {\em Proc. Detection and Classification of Acoustic Scenes and Events
  Workshop {\rm (}DCASE{\rm )}}, pp. 1--5, 2017.

\bibitem{Kim_WASPAA2009_01}
S.~Kim, S.~Narayanan, and S.~Sundaram,
\newblock ``Acoustic topic models for audio information retrieval,''
\newblock {\em Proc. 2009 {IEEE} Workshop on Applications of Signal Processing
  to Audio and Acoustics {\rm (}WASPAA{\rm )}}, pp. 37--40, 2009.

\bibitem{Imoto_MLSP2013_01}
K.~Imoto, Y.~Ohishi, H.~Uematsu, and H.~Ohmuro,
\newblock ``Acoustic scene analysis based on latent acoustic topic and event
  allocation,''
\newblock {\em Proc. IEEE International Workshop on Machine Learning for Signal
  Processing {\rm (}MLSP{\rm )}}, 2013.

\bibitem{Kwon_ISCS2009_01}
H.~Kwon, H.~Krishnamoorthi, V.~Berisha, and A.~Spanias,
\newblock ``A sensor network for real-time acoustic scene analysis,''
\newblock {\em Proc. {IEEE} International Symposium on Circuits and Systems},
  pp. 169--172, 2009.

\bibitem{Giannoulis_EUSIPCO2015_01}
P.~Giannoulis, A.~Brutti, M.~Matassoni, A.~Abad, A.~Katsamanis, M.~Matos,
  G.~Potamianos, and P.~Maragos,
\newblock ``Multi-room speech activity detection using a distributed microphone
  network in domestic environments,''
\newblock {\em Proc. 23rd European Signal Processing Conference {\rm
  (}EUSIPCO{\rm )}}, pp. 1271--1275, 2015.

\bibitem{Dekkers_DCASE2017_01}
G.~Dekkers, S.~Lauwereins, B.~Thoen, M.~W. Adhana, H.~Brouckxon,
  T.~Waterschoot, B.~Vanrumste, M.~Verhelst, and P.~Karsmakers,
\newblock ``The {SINS} database for detection of daily activities in a home
  environment using an acoustic sensor network,''
\newblock {\em Proc. Detection and Classification of Acoustic Scenes and Events
  Workshop {\rm (}DCASE{\rm )}}, pp. 32--36, 2017.

\bibitem{Dekkers_arXiv2018_01}
G.~Dekkers, L.~Vuegen, T.~Waterschoot, B.~Vanrumste, and P.~Karsmakers,
\newblock ``{DCASE} 2018 challenge - task 5: Monitoring of domestic activities
  based on multi-channel acoustics,''
\newblock {\em arXiv preprint arXiv:1807.11246}, 2018.

\bibitem{Tanabe_DCASE2018_01}
R.~Tanabe, T.~Endo, Y.~Nikaido, T.~Ichige, P.~Nguyen, Y.~Kawaguchi, and
  K.~Hamada,
\newblock ``Multichannel acoustic scene classification by blind
  dereverberation, blind source separation, data augmentation, and model
  ensembling,''
\newblock {\em Tech. Rep. DCASE}, pp. 1--4, 2018.

\bibitem{Kurby_DCASE2016_01}
J.~K{\"u}rby, R.~Grzeszick, A.~Plinge, and G.~A. Fink,
\newblock ``Bag-of-features acoustic event detection for sensor networks,''
\newblock {\em Proc. Detection and Classification of Acoustic Scenes and Events
  Workshop {\rm (}DCASE{\rm )}}, pp. 55--59, 2016.

\bibitem{Imoto_EUSIPCO2015_01}
K.~Imoto and N.~Ono,
\newblock ``Spatial-feature-based acoustic scene analysis using distributed
  microphone array,''
\newblock {\em Proc. European Signal Processing Conference {\rm (}EUSIPCO{\rm
  )}}, pp. 739--743, 2015.

\bibitem{Imoto_TASLP2017_01}
K.~Imoto and N.~{Ono},
\newblock ``Spatial cepstrum as a spatial feature using distributed microphone
  array for acoustic scene analysis,''
\newblock {\em {IEEE/ACM} Trans. Audio Speech Lang. Process.}, vol. 25, no. 6,
  pp. 1335--1343, 2017.

\bibitem{Hasegawa_LVAICA2010_01}
K.~Hasegawa, N.~Ono, S.~Miyabe, and S.~Sagayama,
\newblock ``Blind estimation of locations and time offsets for distributed
  recording devices,''
\newblock {\em Proc. Latent Variable Analysis and Signal Separation: 9th
  International Conference, {LVA}/{ICA} 2010}, pp. 57--64, 2010.

\bibitem{Ono_WASPAA2009_01}
N.~Ono, H.~Kohno, and S.~Sagayama,
\newblock ``Blind alignment of asynchronously recorded signals for distributed
  microphone array,''
\newblock {\em Proc. Applications of Signal Processing to Audio and Acoustics
  {\rm (}WASPAA{\rm )}}, pp. 161--164, 2009.

\bibitem{Schmalenstroeer_EUSIPCO2013_01}
J.~Schmalenstroeer and R.~Haeb-Umbach,
\newblock ``Sampling rate synchronization in acoustic sensor networks with a
  pre-trained clock skew error model,''
\newblock {\em Proc. 21st European Signal Processing Conference {\rm
  (}EUSIPCO{\rm )}}, pp. 1--5, 2013.

\bibitem{Miyabe_ESP2015_01}
S.~Miyabe, N.~Ono, and S.~Makino,
\newblock ``Blind compensation of interchannel sampling frequency mismatch for
  ad hoc microphone array based on maximum likelihood estimation,''
\newblock {\em Elsevier Signal Processing}, vol. 107, pp. 185--196, 2 2015.

\bibitem{Shuman_SPM2013_01}
D.~I. Shuman, S.~K. Narang, P.~Frossard, A.~Ortega, and P.~Vandergheynst,
\newblock ``The emerging field of signal processing on graphs: Extending
  high-dimensional data analysis to networks and other irregular domains,''
\newblock {\em {IEEE} Signal Process. Mag.}, vol. 30, no. 3, pp. 83--98, 2013.

\bibitem{Mesaros_DCASE2017_01}
A.~Mesaros, T.~Heittola, A.~Diment, B.~Elizalde, A.~Shah, E.~Vincent, B.~Raj,
  and T.~Virtanen,
\newblock ``{DCASE} 2017 challenge setup: Tasks, datasets and baseline
  system,''
\newblock {\em Proc. Detection and Classification of Acoustic Scenes and Events
  Workshop {\rm (}DCASE{\rm )}}, pp. 85--92, 2017.

\bibitem{Gemmeke_ICASSP2017_01}
J.~F. Gemmeke, D.~P. Ellis, D.~Freedman, A.~Jansen, W.~Lawrence, R.~C. Moore,
  M.~Plakal, and M.~Ritter,
\newblock ``Audio set: An ontology and human-labeled dataset for audio
  events,''
\newblock {\em Proc. {IEEE} International Conference on Acoustics, Speech and
  Signal Processing {\rm (}ICASSP{\rm )}}, pp. 776--780, 2017.

\bibitem{Imoto_EUSIPCO2017_01}
K.~Imoto and N.~Ono,
\newblock ``Acoustic scene classification based on generative model of acoustic
  spatial words for distributed microphone array,''
\newblock {\em Proc. European Signal Processing Conference {\rm (}EUSIPCO{\rm
  )}}, pp. 2343--2347, 2017.

\bibitem{Golub_JHUnivPress1996_01}
G.~Golub and C.~Van Loan,
\newblock {\em Matrix Computations},
\newblock Johns Hopkins University Press, 1996.

\end{thebibliography}
%
\begin{figure}[t]
\begin{center}
\includegraphics[width=0.95\columnwidth]{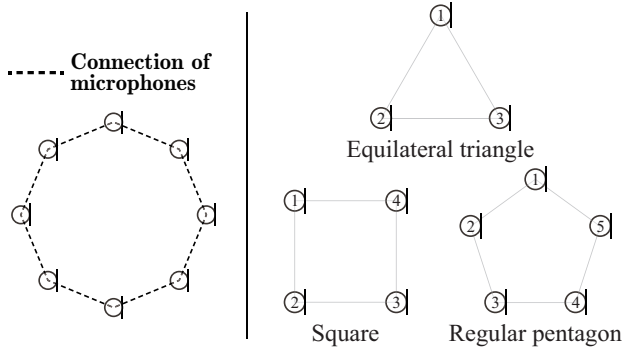}
\end{center}
\caption{Examples of ring graph condition (left) and circularly symmetric microphone arrangements (right)}
\label{fig:symmetric01}
\end{figure}
%
%
\appendix
\section*{\small{Appendix: Graph Cepstrum on Ring Graph}}
\label{sec:Spcase}
As a special case, let us consider a circularly connected condition, namely, the ring graph condition shown in Fig.~\ref{fig:symmetric01}.
Under the ring graph condition, a graph Laplacian is represented as the circulant matrix
%
\begin{align}
\hspace{-10pt} {\bf L}_{\textit{sym}} = \left[ \begin{array}{rrrcrrr}
2&-1&0&\ \ \cdots\!&0&0&-1\ \ \\
-1&2&-1&\ \ \cdots\!&0&0&0\ \ \\
0&-1&2&\ \ \cdots\!&0&0&0\ \ \\
\vdots&\vdots&\vdots&\ \ \ddots\!&\vdots&\vdots&\vdots\ \ \\
0&0&0&\ \ \cdots\!&2&-1&0\ \ \\
0&0&0&\ \ \cdots\!&-1&2&-1\ \ \\
-1&0&0&\ \ \cdots\!&0&-1&2\ \ \end{array} \right]. \hspace{-6pt}
\label{eq:symL}
\end{align}
%
Considering that a circulant matrix is diagonalized by an IDFT matrix ${\bf Z}_{N}$ \cite{Golub_JHUnivPress1996_01} defined by
%
\begin{align}
&{\bf Z}_{N} \! = \! \frac{1}{\sqrt{N}} \! \left[ \begin{array}{cccccc} \!1 \!&\! 1 \!\!&\!\! 1 \!&\!\! \cdots \!\!&\! 1 \!\!&\!\! 1\\
\!1 \!&\! \zeta^{1} \!\!&\!\! \zeta^{2} \!\!&\!\!\! \cdots \!\!\!&\!\! \zeta^{N\!-\!2} \!\!&\!\! \zeta^{N-1}\\
\!1 \!&\! \zeta^{2} \!\!&\!\! \zeta^{4} \!\!&\!\!\! \cdots \!\!\!&\!\! \zeta^{2(N\!-\!2)} \!\!&\!\! \zeta^{2(N\!-\!1)}\\
\!\vdots &\! \vdots \!\!&\!\! \vdots \!\!&\!\!\! \ddots \!\!\!&\!\! \vdots \!\!&\!\! \vdots\\
\!1 \!&\! \zeta^{N\!-\!2} \!\!&\!\! \zeta^{2(N\!-\!2)} \!\!&\!\!\! \cdots \!\!\!&\!\! \zeta^{(N-2)^2} \!\!&\!\! \zeta^{(N\!-\!1)(N\!-\!2)}\\
\!1 \!&\! \zeta^{N\!-\!1} \!\!&\!\! \zeta^{2(N\!-\!1)} \!\!&\!\!\! \cdots \!\!\!&\!\! \zeta^{(N\!-\!2)(N\!-\!1)} \!\!&\!\! \zeta^{(N-1)^2}
\end{array} \!\! \right]\!\!\! 
\end{align}
\begin{align}
\hspace{70pt} \zeta = e^{j2 \pi / N},
\end{align}
%
\noindent the IGFT is identical to the IDFT.
Thus, under the ring graph condition, the GC is identical to the definition of the cepstrum.
Moreover, it is also identical to the SC under the condition of circularly symmetric microphones in an isotropic sound field \cite{Imoto_TASLP2017_01}.
This means that the ring connection in the GC domain corresponds to the circularly symmetric arrangement of microphones in an isotropic sound field.
%
\end{document}